\documentclass[twocolumn,amsmath,amssymb,prc,superscriptaddress,nofootinbib,floatfix,showpacs]{revtex4}
\usepackage{graphicx}
\usepackage{color}
\usepackage{hyperref}
\usepackage{dcolumn}
\usepackage{subfigure}
\usepackage{xspace}

\newcommand{\Dslash}{\ensuremath{D\hspace{-1.5ex} /}}
\newcommand{\Tr}{\ensuremath{\operatorname{Tr}}}
\newcommand{\tr}{\ensuremath{\operatorname{tr}}}

\def\roughly#1{\mathrel{\raise.3ex\hbox{$#1$\kern-.75em%
\lower1ex\hbox{$\sim$}}}}

\newcommand\T{\rule{0pt}{2.6ex}}
\newcommand\B{\rule[-1.2ex]{0pt}{0pt}}

\newcommand{\vev}[1]{\ensuremath{\left\langle #1 \right\rangle}}
\newcommand{\einh}[1]{\ensuremath{\,\text{#1}}}
\newcommand{\MeV}{\einh{MeV}}

\def\Eq#1{Eq.~(\ref{#1})}
\def\Fig#1{Fig.~\ref{#1}}
\def\Tab#1{Tab.~\ref{#1}}

\newcommand{\msig}{\ensuremath{m_{\sigma}}}

\newcommand{\ua}{\ensuremath{U(1)_A}}
\newcommand{\Phibar}{\ensuremath{\bar{\Phi}}}
\newcommand{\LPQM}{\ensuremath{\mathcal{L}_{\textrm{PQM}}}\xspace}

\newcommand{\onefig}{0.8\linewidth}
\newcommand{\twofigs}{0.4\linewidth}

\newcommand{\pT}{\ensuremath{T_0}}
\newcommand{\Tl}{\ensuremath{T_\chi}}
\newcommand{\Ts}{\ensuremath{T_\chi^s}}
\newcommand{\Tchi}{\ensuremath{T_\chi}}
\newcommand{\Td}{\ensuremath{T_d}}

\newcommand{\coloronl}{(color online)\xspace}

\graphicspath{
{./}
{../figures/}
{./figures/}
}
\begin{document}

\title{Thermodynamics of $(2+1)$-flavor QCD: Confronting Models with Lattice Studies}

\author{B.-J. Schaefer}
\email[E-Mail:]{bernd-jochen.schaefer@uni-graz.at}
\affiliation{Institut f\"{u}r Physik, Karl-Franzens-Universit\"{a}t,
  A-8010 Graz, Austria}
\author{M. Wagner}
\email[E-Mail:]{mathias.wagner@physik.tu-darmstadt.de}
\affiliation{Institut f\"{u}r Kernphysik, TU Darmstadt, D-64289
  Darmstadt, Germany}
\affiliation{ExtreMe Matter Institute EMMI,
  GSI Helmholtzzentrum f\"{u}r Schwerionenforschung GmbH,
  D-64291 Darmstadt, Germany}
\author{J. Wambach}
\affiliation{Institut f\"{u}r Kernphysik, TU Darmstadt, D-64289
  Darmstadt, Germany} \affiliation{Gesellschaft f\"{u}r
  Schwerionenforschung GSI, D-64291 Darmstadt, Germany}



\begin{abstract}
  The Polyakov-quark-meson (PQM) model, which combines chiral as well
  as deconfinement aspects of strongly interacting matter is
  introduced for three light quark flavors. An analysis of the chiral
  and deconfinement phase transition of the model and its
  thermodynamics at finite temperatures is given. Three different
  forms of the effective Polyakov loop potential are considered. The
  findings of the $2+1$ flavor model investigations are confronted to
  corresponding recent QCD lattice simulations of the RBC-Bielefeld,
  HotQCD and Wuppertal-Budapest collaborations. The influence of the
  heavier quark masses, which are used in the lattice calculations, is
  taken into account. In the transition region the bulk thermodynamics
  of the PQM model agrees well with the lattice data.
\end{abstract}

\maketitle

\section{Introduction}
\label{sec:intro}

While the underlying microscopic theory of the strong interaction,
Quantum Chromodynamics (QCD), is known for several decades, our
knowledge on its phase structure under extreme conditions is still
rather limited~\cite{MeyerOrtmanns:1996ea,Rischke:2003mt}.
Understanding its critical properties and phase transitions is a major
focus which is addressed in both theoretical and experimental studies.
Heavy-ion collisions in such experiments at the highest currently 
available energies are surprisingly
well-described by ideal relativistic hydrodynamics (see
e.g.~\cite{Kolb:2003dz} and references therein). This demands a
detailed knowledge of the bulk thermodynamics and the equation of
state of strongly-interacting matter.

Different regimes of the QCD phase diagram can be examined by
employing various theoretical methods. For example, lattice QCD
simulations (see e.g.~\cite{FKOP}
) are
applicable to the finite temperature regime of the phase diagram but
at finite chemical potentials the fermion sign problem is still a
considerable obstacle. At finite temperatures recent lattice
calculations describe the QCD thermodynamics reliably since larger
volumes are used and the quark masses are closer to their physical
values. Although most lattice simulations for $(2+1)$-flavor QCD agree
that chiral symmetry is restored by a smooth crossover
transition~\cite{Aoki:2006we} there is still an ongoing discussion
concerning the (pseudo)critical temperatures of the chiral and
deconfinement transition~\cite{Karsch:1998qj}. One group
(RBC-Bielefeld, HotQCD) sees a coincidence of both
transitions~\cite{Cheng:2006qk, Bazavov:2009zn} while the
Wuppertal-Budapest group finds a larger deconfinement temperature
compared to the chiral one by examining the peaks of the chiral and
Polyakov-loop susceptibilities~\cite{Aoki:2006br, Aoki:2009sc}.

Effective NJL-type models can address chiral aspects of the QCD phase
diagram and do not suffer from limitations at finite chemical
potentials. It is expected that the critical behavior of such models
is governed by the same universality class as
QCD~\cite{Pisarski:1983ms}. But, due to the lack of confinement, they
fail in describing the thermodynamics of strongly interacting matter,
in particular the deconfinement transition. However, both transitions
become accessible with extended NJL-type models which are augmented
with the Polyakov loop. Recently, most studies in these Polyakov-loop
NJL-type models (PNJL) focus on two quark flavors with an effective
polynomial or logarithmic Polyakov loop potential~\cite{Megias,
 Ratti:2005jh, Roessner:2006xn,Ghosh:2006qh,
  Sasaki:2006ww, Rossner:2007ik, Blaschke:2007np, Hell:2008cc,
  Abuki:2008nm, Steinheimer:2009hd}. 
   The thermodynamics of these
models has often been compared to older two-flavor lattice data with
large quark masses (e.g.~\cite{AliKhan:2001ek}) while in the model
calculations physical values have been used. Recent $(2+1)$-flavor
studies in the PNJL~\cite{Ciminale:2007sr, Fu:2007xc,
  Fukushima:2008wg, Fukushima:2008is, Fukushima:2009dx} or
Polyakov-quark-meson (PQM) models~\cite{Schaefer:2008ax,
  Schaefer:2009st, Mao:2009aq} indeed agree better with the
corresponding lattice data although the used quark masses still
differ.

Much progress has been achieved in lattice studies for (2+1)-flavor
QCD thermodynamics, e.g.~\cite{Cheng:2006qk, Cheng:2007jq,
  Cheng:2008zh, Bazavov:2009zn, Bernard:2006nj, Aoki:2005vt,
  Aoki:2009sc}. In this work a $(2+1)$-flavor PQM model study is
compared with the most recent lattice data of the HotQCD and
RBC-Bielefeld~\cite{Bazavov:2009zn, Cheng:2008zh} and the
Wuppertal-Budapest (WB) collaborations~\cite{Aoki:2009sc}. In the
lattice simulations of the HotQCD and RBC-Bielefeld collaborations a
ratio of the physical strange quark mass to the light one of 10 is
typically used which yields too large light quark masses. These quark
masses entail a pion mass of $m_\pi \sim 220 \MeV$ and a kaon mass of
$m_K \sim 500 \MeV$ using improved staggered fermions (p4 and asqtad
actions). They observe a coincidence of the chiral and the
deconfinement transition around temperatures $T \sim 185-195 \MeV$.
Coinciding temperatures result also in a two flavor QCD analysis if
quantum fluctuations are incorporated, see e.g.~\cite{Braun:2009gm,
  Schaefer:2007pw}.

For the chiral order parameter and the strange quark number
susceptibility we also compare our results with lattice data from the
Wuppertal-Budapest collaboration which they have obtained for physical
quark masses on $N_\tau=10$ lattices~\cite{Aoki:2009sc}. In contrast
to the other lattice collaborations they find a chiral transition at
temperatures lower than the one for the deconfinement transition,
i.e., $T\chi \sim 146\MeV < T_d \sim 170\MeV$.

The outline of this work is as follows: after introducing the PQM
model for $N_f=2+1$ quark flavors, the implementation of various
effective Polyakov-loop potentials is discussed. The thermodynamic
potential of the PQM model in mean-field approximation is derived in
Sec.~\ref{sec:pot}. The resulting chiral and confinement/deconfinement
phase structure and its dependence on the model parameters and various
Polyakov-loop potentials are presented in Sec.~\ref{sec:phases}. A
detailed comparison for several thermodynamic quantities such as the
equation of state and various susceptibilities with lattice data is
carried out in Sec.~\ref{sec:thermodynamics}. Here, the effect of the
heavier pion masses and different Polyakov loop potential choices is
explicitly considered. Finally, conclusions are drawn in
Sec.~\ref{sec:summary}.

\section{Polyakov-Quark-Meson Model}
\label{sec:pqm}

In order to address the linkage between the chiral symmetry
restoration and aspects of the confinement/deconfinement transition we
introduce the Polyakov-quark-meson (PQM) model for three quark
flavors. It is a straightforward generalization of the two quark
flavor model \cite{Schaefer:2007pw} and is a combination of the chiral
linear $\sigma$-model \cite{Gell-Mann:1960np} with the Polyakov loop
$\Phi(\vec x)$, the thermal expectation value of a color traced Wilson
loop in the temporal direction
\begin{equation}
  \label{eq:defphi} 
  \Phi(\vec x) = \frac{1}{N_c} \langle \ensuremath{\tr}_c \mathcal{P}
  (\vec x) \rangle\ .
\end{equation}
The matrix-valued Polyakov loop operator $\mathcal{P}(\vec x)$ in the
fundamental representation of the $SU(N_c)$ gauge group contains the 
temporal vector field $A_0$,
\begin{equation}
  \label{eq:defP}
  \mathcal{P}(\vec x) = \mathrm{P} \exp \left( i \int_0^{\beta} d\tau A_0
    (\vec x, \tau ) \right)\ ,
\end{equation}
where $\mathrm{P}$ denotes path ordering and $\beta = 1/T$ the inverse
temperature \cite{Polyakov:1978vu}. Since the Polyakov loop in this
model is used as a classical variable, implementation details in
\Eq{eq:defP} are not important in the present work.

The expectation value \Eq{eq:defphi} serves as an order parameter of
the center symmetry in the heavy quark mass limit: it is finite when
the center symmetry is spontaneously broken and vanishes for 
center-symmetric states. Hence, $\Phi$ is finite at high temperatures
corresponding to the deconfined plasma phase and it vanishes at low
temperatures in the confined, center-symmetric phase. However, in a
system with dynamical quarks the center symmetry is always broken but
$\Phi$ still seems to be a good indicator of the
confinement/deconfinement transition. Therefore, we regard $\Phi$ as
an approximate order parameter for this phase transition.

The PQM Lagrangian consists of a quark-meson contribution and a
Polyakov-loop potential $\mathcal{U} (\Phi[A], \Phibar[A])$, which
depends on the Polyakov-loop variable $\Phi$ and its hermitian conjugate
$\Phibar= \langle \ensuremath{\tr}_c \mathcal{P}^\dagger \rangle/
N_c$. The coupling of the uniform temporal background gauge field to
the quarks is achieved by replacing the standard derivative
$\partial_\mu$ in the quark-meson contribution by a covariant
derivative
\begin{equation}
  D_\mu = \partial_\mu - i A_\mu \quad , \quad  A_\mu = \delta_{\mu 0}
  A^0 
\end{equation}
where the $SU(N_c)$ gauge coupling $g_s$ is included in
$A_\mu = g_s A_\mu^a \lambda^a/{2}$. The usual Gell-Mann matrices are
denoted by $\lambda^a$ with $a=1,\ldots, N_c^2-1$. This leads to the
Lagrangian 
\begin{equation}
  \label{eq:lpqm}
  \LPQM = \bar{q}\left(i \Dslash - g \phi_5 \right) q +
  \mathcal{L}_m  
  -\mathcal{U} (\Phi[A], \Phibar[A])\ ,
\end{equation}
where the column vector $q=(u, d, s)$ denotes the quark field for
$N_f = 3$ flavors and $N_c=3$ color degrees of freedom. 
For three quark flavors the interaction between the quarks and the
meson nonets is implemented by a flavor-blind Yukawa coupling $g$ and
the meson matrix
\begin{equation}
  \phi_5 = T_a\left( \sigma_a + i \gamma_5 \pi_a
  \right) \ ,
\end{equation}
where the $T_a = \lambda_a/2$, $a=0,\ldots, 8$ are the nine
generators of the $U(3)$ symmetry with
$\lambda_0 = \sqrt{\frac{ 2}{3}}\ \bf 1$. The generators $T_a$ are
normalized to $\Tr (T_a T_b) = \delta_{ab}/2 $ and obey the $U(3)$
algebra $[T_a,T_b]=if_{abc}T_c$ and $\{T_a,T_b \}=d_{abc}T_c$
respectively with the corresponding standard symmetric $d_{abc}$ and
antisymmetric $f_{abc}$ structure constants of the $SU(3)$ group and
\begin{eqnarray}
f_{ab0} = 0 &,& d_{ab0} = \sqrt{\frac{ 2}{3}} \delta_{ab}\ .
\end{eqnarray}
The nine scalar ($J^P = 0^+$) mesons are labeled by the $\sigma_a$
fields and accordingly the nine pseudoscalar ($J^P = 0^-$) mesons by
the $\pi_a$ fields.

The remaining, purely mesonic contribution reads
\begin{eqnarray}
\label{eq:mesonL}
  \mathcal{L}_m &=& \Tr \left( \partial_\mu \phi^\dagger \partial^\mu
    \phi \right)
  - m^2 \Tr ( \phi^\dagger \phi) -\lambda_1 \left[\Tr (\phi^\dagger
    \phi)\right]^2 \nonumber \\
  &&  - \lambda_2 \Tr\left(\phi^\dagger \phi\right)^2
  +c   \left(\det (\phi) + \det (\phi^\dagger) \right)\nonumber \\
  && + \Tr\left[H(\phi+\phi^\dagger)\right]\ ,
\end{eqnarray}
with the abbreviation 
\begin{equation}
  \label{eq:fields}
  \phi = T_a \phi_a = T_a \left(\sigma_a + i \pi_a\right)\ ,
\end{equation}
representing a complex $(3\times 3)$-matrix for $N_f=3$.

Chiral symmetry is broken explicitly by the last term in
\Eq{eq:mesonL} where 
\begin{equation}
  H = T_a h_a
\end{equation}
is a again $(3\times 3)$-matrix with nine external parameters $h_a$.
The $\ua$-symmetry is explicitly broken by the 't Hooft determinant
term with a constant strength $c$. Further details concerning the
three-flavor quark-meson part of the PQM model can be found in
\cite{Schaefer:2008hk}.
 
The final form of the effective gluon field potential $\mathcal{U}$ in
\Eq{eq:lpqm} is constructed in terms of the Polyakov loop variables
$\Phi$ and $\Phibar$. It preserves the center symmetry of the pure
Yang-Mills (YM) theory \cite{Pisarski:2000eq, Dumitru:2001xa}.
However, several explicit choices of the Polyakov loop potential are
possible which we collect in the following.

\subsection{Polyakov loop potentials}
\label{sec:polypots}

The functional form of the Polyakov loop potential is motivated by the
underlying QCD symmetries in the pure gauge limit, i.e. for infinitely 
heavy quarks. The different parameterizations reproduce a
first-order transition at $T\sim 270\MeV$ in the pure gauge sector for
$N_c=3$ \cite{Ratti:2005jh, Roessner:2006xn, Fukushima:2008wg}.

The simplest choice is based on a Ginzburg-Landau ansatz
\cite{Ratti:2005jh, Schaefer:2007pw}. The underlying $Z(3)$ center
symmetry is broken spontaneously in the pure Yang-Mills case which
should be respected by the ansatz. Hence, an expansion in terms of the
order parameter results in
\begin{multline}
\label{eq:upoly}
  \frac{\mathcal{U}_{\text{poly}}}{T^{4}}= -\frac{b_2}{4}
\left(|\Phi|^2+|\bar\Phi|^2 \right) \\ 
-\frac{b_3}{6}(\Phi^3+\Phibar^3)+\frac{b_4}{16}
\left(|\Phi|^2+|\Phibar|^2\right)^2
\end{multline}
with a temperature-dependent coefficient
\begin{equation}
  b_2(T) = a_0 + a_1 \left(\frac{T_0}{T}\right) + a_2
  \left(\frac{T_0}{T}\right)^2 + a_3 \left(\frac{T_0}{T}\right)^3\ . 
\end{equation}
The cubic $\Phi$ terms in \Eq{eq:upoly} are required to break the
$U(1)$ symmetry of the remaining terms down to $Z(3)$. The potential
parameters are adjusted to the pure gauge lattice data such that the
equation of state and the Polyakov loop expectation values are
reproduced. This yields the following set of
parameters~\cite{Ratti:2005jh}:
\begin{gather*}
a_0=6.75\ , \quad a_1=-1.95\ , \quad a_2=2.625\ , \quad a_3 = -7.44\\
 b_3=0.75\ , \quad b_4=7.5\ .
\end{gather*}
The parameter $T_0 = 270 \MeV$ corresponds to the transition
temperature in the pure YM theory. The extension of the potential
\Eq{eq:upoly} to finite chemical potential and the remaining ambiguity
is discussed in \cite{Schaefer:2007pw}. At vanishing chemical
potential the constraint $\bar\Phi = \Phi^\dagger$ is still valid and
the potential corresponds to a polynomial expansion in $\Phi$ and
$\bar\Phi$ ,including the product term $\Phi\bar\Phi$ as in
Refs.~\cite{Pisarski:2000eq, Ratti}.

An improved ansatz 
\cite{Roessner:2006xn} results in
\begin{multline}
\label{eq:ulog}
\frac{\mathcal{U}_{\text{log}}}{T^{4}}= -\frac{1}{2}a(T) \Phibar \Phi \\
+ b(T) \ln \left[1-6 \Phibar\Phi + 4\left(\Phi^{3}+\Phibar^{3}\right)
  - 3 \left(\Phibar \Phi\right)^{2}\right]
\end{multline}
with the temperature-dependent coefficients
\begin{equation}
  a(T) =  a_0 + a_1 \left(\frac{T_0}{T}\right) + a_2 \left(\frac{T_0}{T}\right)^2
\end{equation}
and
\begin{equation}
  b(T) = b_3 \left(\frac{T_0}{T}\right)^3\ .
\end{equation}
Here, the logarithmic form constrains $\Phi$ and $\bar\Phi$ to values
smaller than one, in contrast to the polynomial potential
ansatz.

Again, the parameters of \Eq{eq:ulog} are fitted to the pure gauge
lattice data, resulting in~\cite{Roessner:2006xn}
\begin{equation}
a_0 = 3.51\ , \quad a_1= -2.47\ , \quad a_2 = 15.2\ ,\quad b_3=-1.75\ 
\end{equation}
Both fits reproduce equally well the equation of state and the
Polyakov loop expectation value.

A third version, proposed by Fukushima~\cite{Fukushima:2008wg}, is
inspired by a strong-coupling analysis
\begin{multline}
\label{eq:ufuku}
\frac{\mathcal{U}_{\text{Fuku}}}{T^4} = -\frac{b}{T^3}  \left[54 e^{-a/T} \Phi \Phibar  \right.\\
\left.+ \ln \left(1 - 6 \Phi \Phibar - 3 (\Phi \Phibar)^2 + 4 (\Phi^3
    + \Phibar^3)\right)\right]
\end{multline}
and has only two parameters $a$ and $b$. The first term is 
reminiscent of the nearest-neighbor interaction and the logarithmic
term is again due to the Haar measure as in (\ref{eq:ulog}). The
parameter $a$ determines the deconfinement transition, i.e., the
transition temperature in pure gauge theory, while $b$ controls the
mixing of the chiral and the deconfinement transition. In this
version, the parameters are not fitted to lattice data but they 
reproduce a first-order transition at $T_0 \sim 270 \MeV$ in the pure
gauge sector for
\begin{equation}
  a=664 \MeV\ , \quad b =  (196.2 \MeV)^3\ \ .
\end{equation}
In a mean-field PNJL analysis with this potential ansatz a coincidence
of the chiral and deconfinement transitions at $T_c\sim200 \MeV$ for
vanishing chemical potential is found~\cite{Fukushima:2008wg}.

Another significant difference of the last potential ansatz is visible
in the thermodynamic pressure at high temperatures. In this case, both
the unconfined transverse gluons as well as the Polyakov loop which
corresponds to longitudinal gluons contribute to the pressure. But
since the Polyakov loop potentials are fitted to pure gauge lattice
data, they thus reproduce the pressure of both the longitudinal and
the transverse gluons and might therefore overcount the degrees of
freedom in the chirally symmetric phase. However, the potential ansatz
by Fukushima excludes these transverse gluon contributions at high
temperatures. This difference will become important when thermodynamic
quantities, obtained with various Polyakov loop potentials, are compared
to lattice data. However, for temperatures $T \lesssim 1.5 T_0$ the
deviations of the Polyakov loop potentials to the pressure are
negligible.

\begin{table}[thbp]
\centering
\begin{tabular}{|c||c|c|c|c|c|c|}
\hline $N_f$ & 0 & 1 & 2 &2+1 & 3\\\hline
$\pT [\MeV]$ & 270 & 240 & 208 & 187 & 178\\\hline
\end{tabular}
\caption{\label{tab:ptNf}The 
  critical temperature $\pT$ for $N_f$ massless flavors according
  to~\cite{Schaefer:2007pw}. The value for $2+1$ flavors has been
  estimated by using HTL/HDL theory for a massive strange quark with
  $m_s=150 \MeV$.}  
\end{table}

In the presence of dynamical quarks, the running coupling of QCD is
modified by fermionic contributions. The size of this effect can 
be estimated within perturbation theory, see e.g. 
\cite{Braun}
This leads to an $N_f$-dependent modification of the expansion
coefficients in the polynomial Polyakov loop potential. The effect 
can be mapped onto an $N_f$-dependent $\pT$
resulting in Tab.~\ref{tab:ptNf} \cite{Schaefer:2007pw}. Similar to
\cite{Braun}, 
the critical temperature $\pT$ decreases for increasing $N_f$.

In this work the $\pT$-dependence on the phase structure will also be
investigated for the logarithmic and polynomial potential.

\section{Thermodynamic Potential}
\label{sec:pot}

In order to investigate the phase structure of the PQM model the
thermodynamic potential is evaluated in mean-field approximation
similar to \cite{Schaefer:2007pw, Schaefer:2008hk}.
In the derivation we will consider the dependence on the temperature
and three quark chemical potentials, $\mu_f$, one for each flavor $f$.
However, in the isospin-symmetric case the light masses are
degenerated, i.e. $m_l \equiv m_u=m_d$ and the thermodynamic potential
depends only on two independent quark chemical potentials,
$\mu_l = (\mu_u+\mu_d)/2$ and $\mu_s$. {Note, that the quark chemical
  potential $\mu_q$ is simply related to the baryochemical potential
  $\mu_B$ by $\mu_q = \mu_B/3$ with $\mu_B = 2\mu_l + \mu_s$.} Thus,
only two order parameters $\sigma_0$ and $\sigma_8$ emerge in the
singlet-octet basis. In the following these condensates are rotated to
the non-strange ($\sigma_x$) and strange ($\sigma_y$) basis, see
\cite{Schaefer:2008hk} for details.

In mean-field approximation the thermodynamic potential consists of
the mesonic $U\left(\sigma_{x},\sigma_{y}\right)$, the quark/antiquark
$\Omega_{\bar{q}{q}}$ and the Polyakov loop contributions
\begin{equation}
  \label{eq:grandpot}
  \Omega = U \left(\sigma_{x},\sigma_{y}\right) +
  \Omega_{\bar{q}{q}} \left(\sigma_{x},\sigma_{y}, \Phi,\Phibar \right) +
  \mathcal{U}\left(\Phi,\Phibar\right)\ . 
\end{equation}
Explicitly, the mesonic contribution reads
\begin{multline}\label{eq:umeson}
  U(\sigma_{x},\sigma_{y}) = \frac{m^{2}}{2}\left(\sigma_{x}^{2} +
  \sigma_{y}^{2}\right) -h_{x} \sigma_{x} -h_{y} \sigma_{y}
 - \frac{c}{2 \sqrt{2}} \sigma_{x}^2 \sigma_{y}
\\
  + \frac{\lambda_{1}}{2} \sigma_{x}^{2} \sigma_{y}^{2}+
  \frac{1}{8}\left(2 \lambda_{1} +
    \lambda_{2}\right)\sigma_{x}^{4}+\frac{1}{8}\left(2 \lambda_{1} +
    2\lambda_{2}\right) \sigma_{y}^{4}\ ,
\end{multline}
and has six parameters
$m^{2}, c, \lambda_{1}, \lambda_{2}, h_{x}, h_{y}$. They are fitted to
well-known pseudoscalar meson masses
$m_{\pi}, m_{K}, m^{2}_{\eta}+m^{2}_{\eta'}$ and two weak-decay constants
$f_{\pi},f_{K}$. The mass of the scalar $\sigma$ meson, $\msig$, is
used to complete the fit. Since the experimental situation concerning
this resonance, cf.~\cite{Yao:2006px} is not clear we vary $\msig$
from 500 MeV to 800 MeV which results in different parameter sets, see
also \cite{Schaefer:2008hk}.

The Polyakov loop variables are coupled to the fermionic part via
\begin{widetext}
  \begin{multline} \label{eq:Omegaqq} \Omega_{\bar qq}( \sigma_{x} ,
    \sigma_{y}, \Phi,\Phibar ) = -2T \sum_{f=u,d,s} \int\frac{d^3p}{(2\pi)^3}
    \left\{\ln \left[1 + 3 (\Phi + \bar \Phi
        e^{-(E_{q,f}-\mu_{f})/T})e^{-(E_{q,f}-\mu_{f})/T} +
        e^{-3(E_{q,f}-\mu_{f})/T}\right]\right.\\
    + \left. \ln \left[1 + 3 (\bar \Phi + \Phi
        e^{-(E_{q,f}+\mu_{f})/T})e^{-(E_{q,f}+\mu_{f})/T} +
        e^{-3(E_{q,f}+\mu_{f})/T}\right] \right\}\ .
\end{multline}
\end{widetext}
The flavor-dependent single-particle energies are given by
\begin{equation}
  E_{q,f}= \sqrt{k^2 + m_f^2}
\end{equation}
with the flavor-dependent quark masses
\begin{align}\label{eq:qmasses}
  m_l = g \sigma_x /2 &\quad\text{and}\quad m_s = g \sigma_y /\sqrt{2}
\end{align}
for the light and strange quarks, respectively. The Yukawa coupling
$g$ is fixed to reproduce a light constituent quark mass of
$m_{q} \approx 300\MeV$. This in turn predicts a strange constituent
quark mass of $m_s \approx 433$ MeV. From \Eq{eq:Omegaqq} one
recognizes the suppressions of single-quark contributions in the
hadronic phase where $\bar\Phi, \Phi \sim 0$ while baryon-like
objects, which are composites of $N_c=3$ quarks, contribute to the
potential. In this way some confinement properties are mimicked in
this model in a statistical sense. On the other hand, for
$\bar\Phi, \Phi \sim 1$, the Polyakov loop modifications decouple from
the fermionic part and \Eq{eq:Omegaqq} turns to the usual
quark/antiquark contribution of the pure quark-meson model
\cite{Schaefer:2008hk}.

Finally, the temperature- and quark chemical potential dependence of
the four order parameters for the chiral and deconfinement transition
are determined as solutions of the corresponding equations of motion.
These coupled equations are obtained by minimizing the grand
potential, \Eq{eq:grandpot}, with respect to the four constant mean-fields $\vev\sigma_x$, $\vev\sigma_y$, $\vev\Phi$ and $\vev{\bar\Phi}$:
\begin{equation}
  \label{eq:pqmeom}
  \left.\frac{ \partial \Omega}{\partial 
      \sigma_x} = \frac{ \partial \Omega}{\partial \sigma_y}  = \frac{
      \partial \Omega}{\partial \Phi}  =\frac{ \partial
      \Omega}{\partial \Phibar} 
  \right|_{\text{min}} = 0\ ,
\end{equation}
where
$\text{min}=\left(\sigma_x=\vev{\sigma_x}, \sigma_y=\vev{\sigma_y},
\Phi=\vev{\Phi}, \bar\Phi=\vev{\bar\Phi} \right)$ labels the global
minimum.

\section{Finite-temperature phase transitions}
\label{sec:phases}

The (pseudo)critical temperatures for the various transitions are
determined as the inflection point of the corresponding order
parameters. The resulting critical temperatures at vanishing chemical
potentials for the non-strange $\Tl$ and strange chiral transitions
$\Ts$ as well as for the deconfinement transition $\Td$ for three
different Polyakov loop potentials and various combinations of
$m_\sigma$ and $T_0$ are summarized in Tab.~\ref{tab:t0pqm}.

In general, the inclusion of the Polyakov loop shifts the
(non-strange) chiral transition at vanishing quark chemical potential,
$\mu_q=0$, to higher temperatures. In addition, $m_\sigma$ and $T_0$
affect the chiral and deconfinement transition temperatures and their
interplay as follows: larger values of $m_\sigma$ and $T_0$ shift
again the transition to higher temperatures. A similar trend
concerning the $m_\sigma$-sensitivity on the phase structure was
observed in a pure three flavor quark-meson model
\cite{Schaefer:2008hk}.

For the chosen parameter sets we find $\Tl \geq \Td$. This behavior is
expected since the Polyakov loop suppresses fermionic contributions to
the thermodynamic potential in the confined phase. However, these
contributions drive the chiral phase transition and thus a delayed
deconfinement transition always delays the chiral symmetry
restoration. This becomes most obvious for small values of
$m_\sigma \approx 500-600 \MeV$, where the chiral transition occurs in
the pure QM model already at $\Tl \lesssim 150 \MeV$. As a
consequence, the PQM model cannot reproduce the Wuppertal-Budapest
scenario~\cite{Aoki:2006br, Aoki:2009sc} with
$\Tl \ll \Td \approx \Ts$. Only for the polynomial Polyakov loop
potential with the extreme parameters $T_0=270\MeV$ and
$m_\sigma=500\MeV$ the case $\Tl < \Td$ is seen. However, the
deconfinement transition is a very smooth crossover and overlaps with
the chiral transition.

For the logarithmic and Fukushima potential only few parameter sets
($m_\sigma=500\MeV$ and $600\MeV$) are found where both transitions
coincide. Even in the deconfined phase the quark contribution to the
thermodynamic potential is not sufficiently large to restore chiral
symmetry immediately and thus $\Tl > \Td$ is obtained.

\begin{table}[htbp]
\centering
\newcolumntype{C}{>{$\,}c<{\,$}}
\begin{tabular}{|cCC || C | C |C|c|}
\hline
\T \B $\mathcal{U}$ & T_0  & m_\sigma & \Tl & \Ts & \Td &  $\Tl \approx \Td $\\
\hline \hline
 - & - & 800 & 184 & 278 &- & \\
 - & - & 600 & 146 & 248 &- & \\
 - & - & 500 & 129 & 238 &- & \\
\hline \hline 
poly & 270 & 800 & 228 & 306 & 227 & $\star$ \\
poly & 208 & 800 & 204 & 297 & 178 &  \\
poly & 187 & 800 & 197 & 295 & 161 &  \\
poly & 178 & 800 & 194 & 273 & 154 &  \\\hline
poly & 270 & 600 & 204 & 262 & 204 & $\star$ \\
poly & 208 & 600 & 179 & 270 & 179 & $\star$ \\
poly & 187 & 600 & 171 & 267 & 171 & $\star$ \\
poly & 178 & 600 & 166 & 236 & 166 & $\star$ \\\hline
poly & 270 & 500 & 193 & 254 & 220 & \\
poly & 208 & 500 & 168 & 239 & 168 & $\star$\\
poly & 187 & 500 & 159 & 237 & 159 & $\star$\\
poly & 178 & 500 & 156 & 236 & 156 & $\star$\\
\hline\hline
log & 270 &  800 & 228 & 302 & 208 & \\
log & 208 &  800 & 208 & 293 & 166 & \\
log & 187 &  800 & 204 & 251 & 150 & \\\hline
log & 270 &  600 & 207 & 274 & 207 & $\star$\\
log & 208 &  600 & 180 & 286 & 165 & \\
log & 187 &  600 & 172 & 283 & 150 & \\\hline
log & 270 &  500 & 198 & 264 & 198 & $\star$\\
log & 208 &  500 & 168 & 255 & 168 & $\star$\\
log & 187 &  500 & 159 & 252 & 150 & \\\hline \hline
Fuku & - & 800 & 214 & 289 & 193 & \\
Fuku & - & 600 & 189 & 260 & 189 & $\star$\\
Fuku & - & 500 & 179 & 250 & 179 & $\star$\\\hline
\end{tabular}
\caption{The (pseudo)critical temperatures of the chiral transition for the light ($\Tl$),
  strange ($\Ts$) and the deconfinement transition ($\Td$) at
  $\mu_q=0$ for different $\msig$ and Polyakov loop potentials in units of MeV,
  evaluated for physical pion masses. In the last column the
  coincidence of the chiral and deconfinement transition is labeled
  with  $\star$.} 
  \label{tab:t0pqm}
\end{table}

In the following we choose a parameter set where both transitions
coincide at $\mu_q=0$ as suggested by the RBC-Bielefeld and HotQCD
lattice calculations~\cite{Cheng:2006qk, Bazavov:2009zn}. This is the
case for $m_\sigma=600\MeV$ and $T_0=270\MeV$ where all three Polyakov
loop potentials yield coinciding transitions. This choice also
excludes modifications caused by different values of $m_\sigma$.
Moreover, recent studies \cite{CapYnd}
point
to $m_\sigma$-values which lie in the lower range of the broad
experimental $\sigma$-resonance region $(400 \MeV\ldots
1200\MeV)$~\cite{Yao:2006px}.

\begin{figure*}[htbp]
  \centering \subfigure[$\ $Polyakov loop expectation value
  $\vev \Phi$]{\label{sfig:pqmpoly}
    \includegraphics[width=\twofigs]{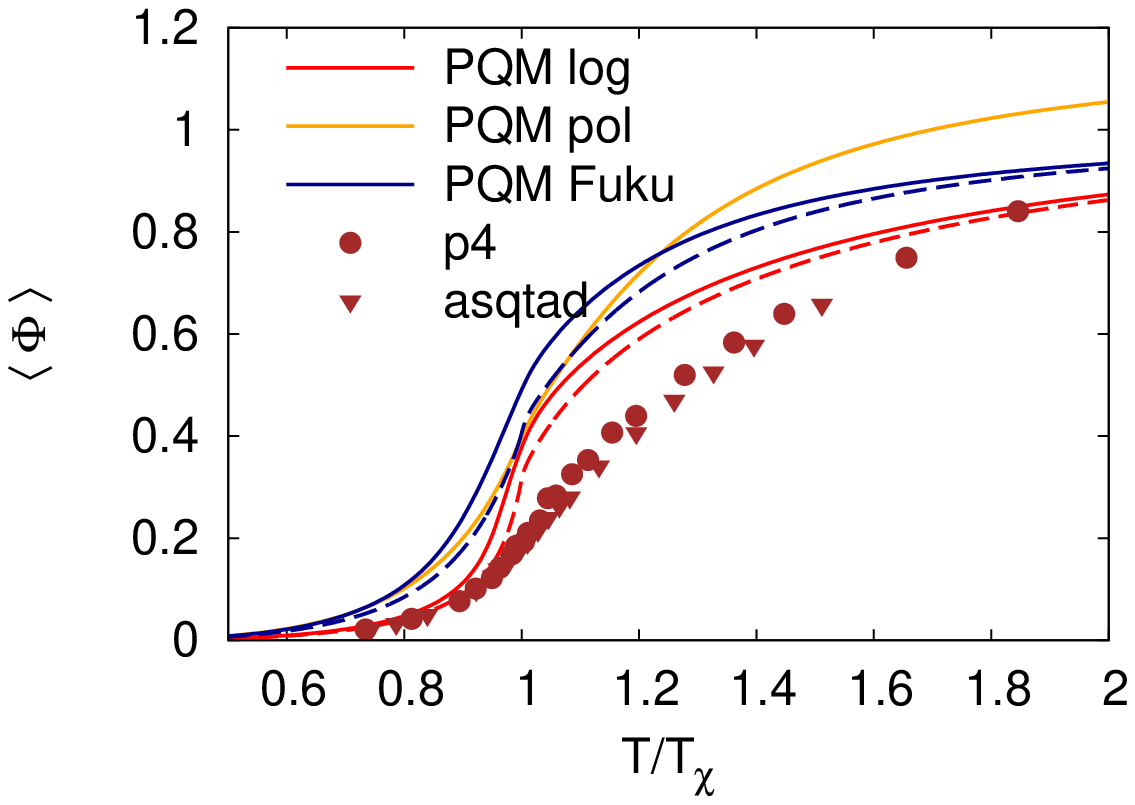}}
    \subfigure[$\ $Subtracted condensate $\Delta_{l,s}$]{\label{sfig:pqmdeltals}
    \includegraphics[width=\twofigs]{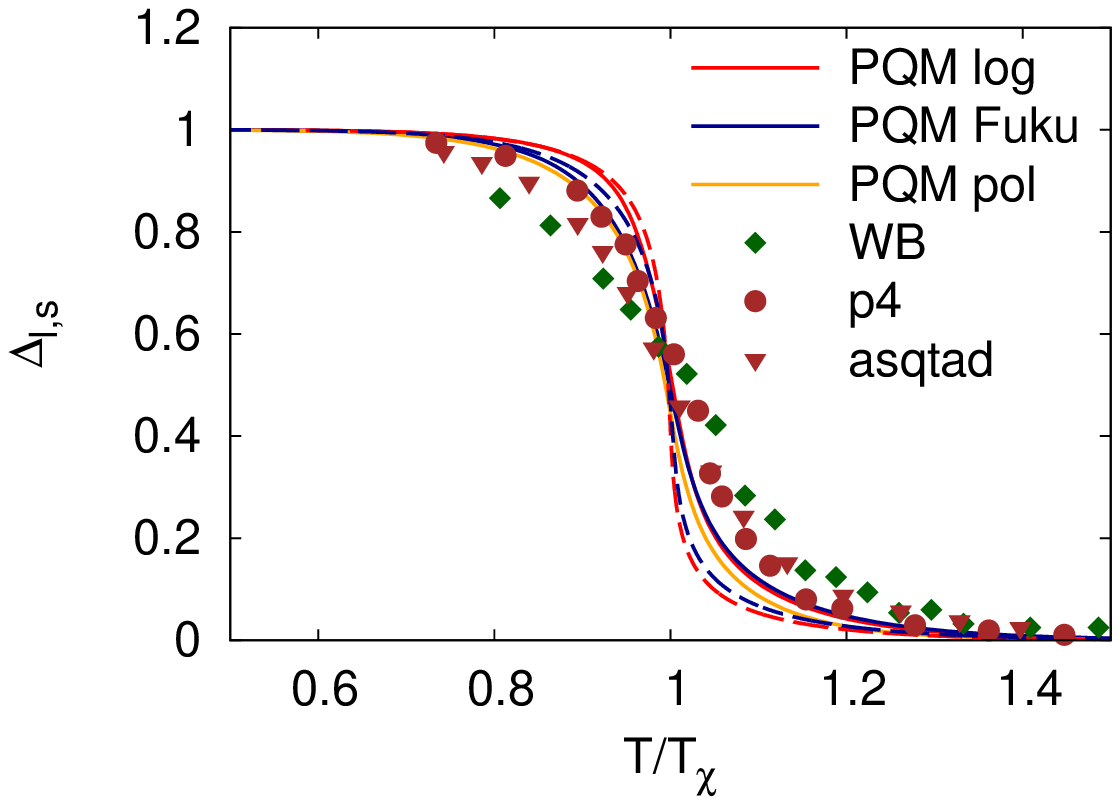}}
  \caption{\label{fig:pqmpolsigmpi}\coloronl The Polyakov loop
    expectation value $\vev \Phi$ (left panel) and the subtracted
    condensate $\Delta_{l,s}$ (right panel) as a function of
    temperature for different Polyakov loop potentials in comparison
    to lattice data of the HotQCD (p4 and asqtad,
    $N_\tau=8$)~\cite{Bazavov:2009zn} and Wuppertal-Budapest
    collaborations (WB, $N_\tau=10,
    m_\pi=135\MeV$)~\cite{Aoki:2009sc}. Solid lines correspond to
    larger pion and kaon masses as used in the HotQCD simulations,
    dashed lines to physical masses.}
    
\end{figure*}

For a proper comparison of the model results with the
HotQCD~\cite{Bazavov:2009zn} and RBC-Bielefeld lattice
data~\cite{Cheng:2006qk, Cheng:2007jq, Cheng:2008zh}, we also adjust
the pion masses accordingly. We denote the mass values of
$m_K = 503\MeV$ and $m_\pi=220 \MeV$ which are used in lattice
simulations as `lattice point' and the measured values of
$m_K = 495\MeV$ and $m_\pi=138 \MeV$ as the 'physical point'. In the
model calculation the tuning of the meson masses is achieved by
adjusting the explicit symmetry breaking parameters $h_x$ and $h_y$
accordingly. Heavier meson masses yield also slightly heavier
constituent quark masses
\begin{equation*}
m_l \approx 322 \MeV\, (300 \MeV)\ ,\  m_s \approx 438\, \MeV (433 \MeV)
\end{equation*}
where the values in parentheses are obtained at the physical point.

The increased meson masses shift the transition to higher
temperatures, see \Tab{tab:t0pqmlat}. The strongest shift is seen in
the non-strange chiral transition since the change of the light quark
masses is larger than that of the strange quark mass in this sector. Both,
the non-strange and the deconfinement transition still coincide.
The strange quark sector is almost unaffected.

\begin{table}[b]
\centering
\newcolumntype{C}{>{$\,}c<{\,$}}
\begin{tabular}{|cCC || C | C |C|c|}
\hline
\T \B $\mathcal{U}$ & T_0  & m_\sigma & \Tl & \Ts & \Td &  $\Tl \approx \Td $\\
\hline \hline
 - & - & 600 & 164 & 250 &- & \\
\hline 
poly & 270 & 600 & 216 & 263 & 215 & $\star$ \\
log & 270 &  600 & 215 & 275 & 208 & $\star$\\
Fuku & - & 600 & 200 & 261 & 196 & $\star$\\\hline
\end{tabular}
\caption{Pseudocritical temperatures similar to \Tab{tab:t0pqm} but
  for $m_\pi=220\MeV$. See text for details.} 
  \label{tab:t0pqmlat}
\end{table}

In \Fig{sfig:pqmpoly} the Polyakov loop expectation value $\vev \Phi$
is shown as a function of temperature for three different Polyakov
loop potentials. The solid lines correspond to heavier meson masses
evaluated at the lattice point and the dashed lines to the physical
ones. The expectation values are slightly larger for Fukushima's
potential than for the logarithmic potential which exhibits a sharper
crossover. The influence of the higher meson masses is mild. The
observed shift may be explained by the larger $\Tchi/T_0$ ratio for
heavier pion masses. The lattice data show a rather smooth transition
with significant lower values for the Polyakov loop expectation value.
For clarity we omit the error bars of the lattice data here and in the
following figures.

The sharper transition in the model as well as the higher values of
the Polyakov loop are probably a remnant of the construction of the
Polyakov loop potentials in the pure gauge sector, where a strong
first order transition is obtained at $T\sim270\MeV$. In the broken
phase, the logarithmic potential comes closest to the lattice data,
while the difference between the model results and the lattice is
considerable in the restored phase for all potentials.

Only for the polynomial potential the expectation value increases
above unity which hampers its interpretation as an order parameter,
related to the free energy of a static color charge. However, these
findings are in agreement with other model studies for three quark
flavors, e.g.~\cite{Mao:2009aq, Fu:2007xc}. In a previous two-flavor
PQM study $\vev \Phi>1$ was only observed at non-vanishing chemical
potential~\cite{Schaefer:2007pw}. For the other Polyakov loop
potentials the logarithmic divergence avoids expectation values larger
than one.

In contrast to the Polyakov loop expectation value, a comparison of
the chiral order parameter $\vev{\bar{q}q}$ is more involved since
some renormalization/normalization factors are unknown. Thus, a more
suitable quantity, to compare with, is the ratio

\begin{equation}
  \Delta_{l,s} = \frac{\langle \bar{l}l\rangle(T) -
    (\hat{m}_l / \hat{m}_s)\langle \bar{s}s\rangle(T)}{\langle
    \bar{l}l\rangle(0) - (\hat{m}_l / \hat{m}_s)\langle
    \bar{s}s\rangle(0)} 
\end{equation}
which includes the non-strange $\langle \bar{l}l\rangle$ and strange
$\langle \bar{s}s\rangle$ condensate. This corresponds to
\begin{equation}
  \Delta_{l,s} = \frac{\vev{\sigma_{x}}(T) -
    ({h_x}/{h_y})\vev{\sigma_{y}}(T)}{\vev{\sigma_{x}}(0) -
    ({h_x}/{h_y})\vev{\sigma_{y}}(0)} 
\end{equation}
in our model calculation, where the quark condensates have been
replaced by $\vev {\sigma_{x}}$ and $\vev{ \sigma_y}$ accordingly.
Since the bare quark masses $\hat{m}_l, \hat{m}_s$ are not directly
available, the ratio of the explicit symmetry breaking parameters
$h_{x,y}$ is used instead. They are proportional to the bare quark
masses up to a constant that drops out in the ratio, see
e.g.~\cite{MeyerOrtmanns:1996ea}.

\begin{figure*}[htbp!]
\centering
\subfigure[$\ $ Normalized pressure $p/p_{SB}$] {\label{sfig:pressure}
  \includegraphics[width=\twofigs]{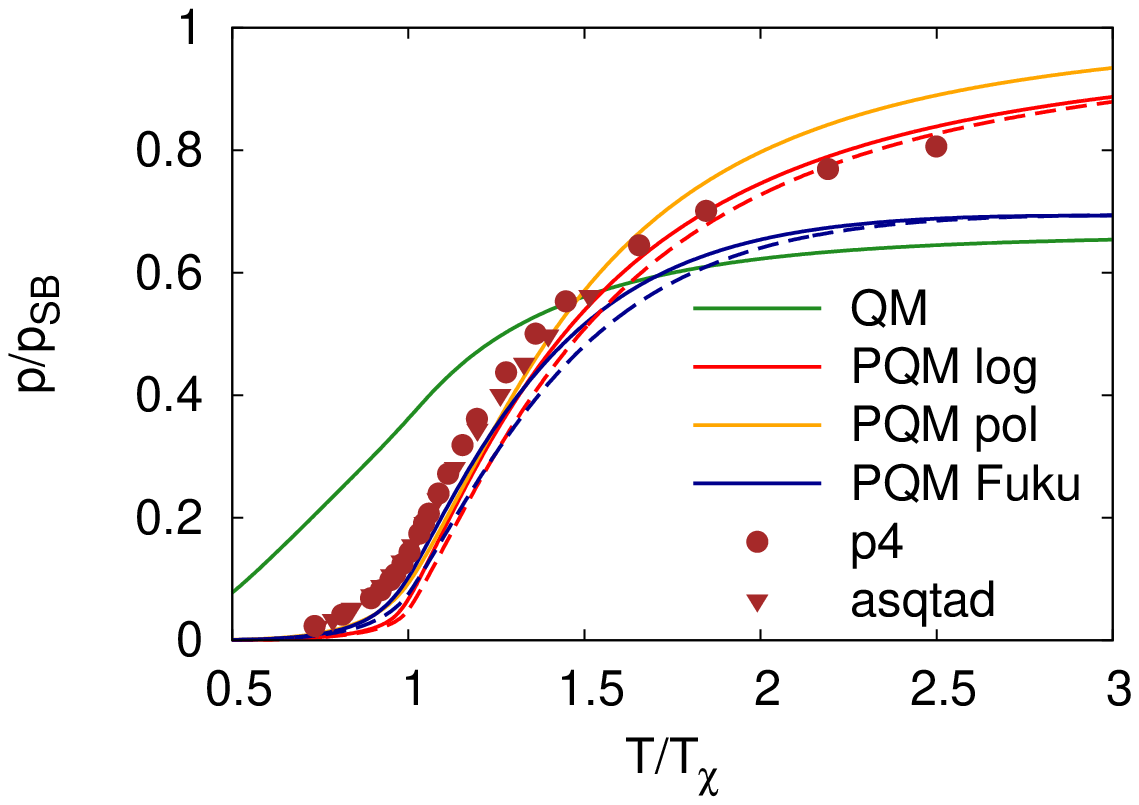}}
\subfigure[$\ $ Interaction measure $\Delta/T^4$]
{\label{sfig:e3p}
  \includegraphics[width=\twofigs]{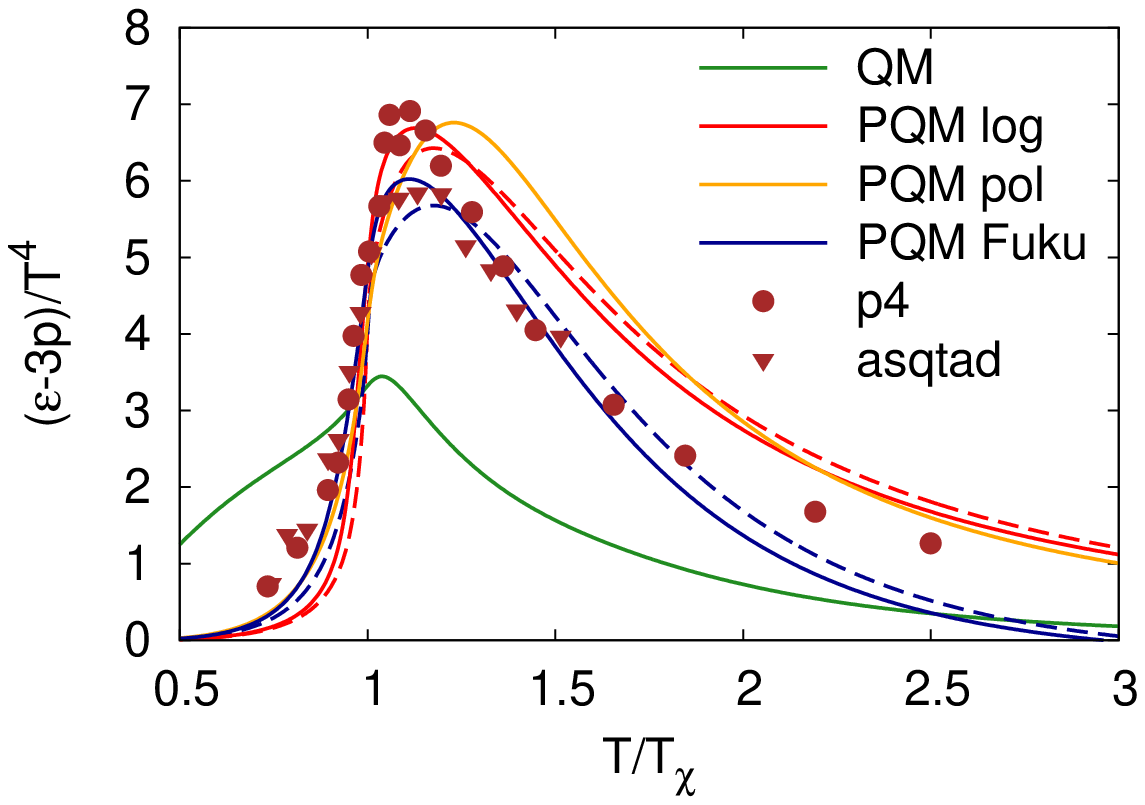}}
\caption{\label{fig:pqmeos}\coloronl The normalized pressure (left
  panel) and the interaction measure (right panel) as a function of
  temperature. The model calculations (PQM model with various Polyakov
  loop potentials and the QM model) are compared to lattice data
  ($N_\tau=8$, p4 and asqtad actions) from ~\cite{Bazavov:2009zn}.
  Solid lines correspond to larger pion and kaon masses as used in the
  lattice simulations, dashed lines to physical masses.}
\end{figure*}

In \Fig{sfig:pqmdeltals} the results for the subtracted condensate
$\Delta_{l,s}$ are shown. The model calculations for $\Delta_{l,s}$
are in better agreement with the lattice data than the ones for the
Polyakov loop expectation value. The lattice data exhibit a somewhat
smoother transition than the model results. The logarithmic Polyakov
loop potential generates the sharpest transition while Fukushima's and
the polynomial potential are closer to the lattice results. The
influence of the larger meson masses is more pronounced in the broken
phase, where the transition becomes smoother for larger meson masses.
The data of the Wuppertal-Budapest group suggest a much smoother
transition. This effect is enhanced in the figure due to the
transition temperature normalization. The transition temperature is
much smaller in the Wuppertal-Budapest simulation than for the model
or HotQCD simulations. Since the Wuppertal-Budapest data are obtained
for physical quark masses they have to be compared to the dashed model
curves which suggest an even sharper transition than the ones obtained
for heavier pion masses.

\section{QCD Thermodynamics}
\label{sec:thermodynamics}

In order to address the modifications of the Polyakov loop on the
equilibrium thermodynamics we evaluate several thermodynamic
quantities and compare the PQM model analysis with the QM model
without the Polyakov loop and recent QCD lattice simulations. The bulk
thermodynamic observables such as pressure, energy and entropy
density are sensitive to the change from hadronic to quark-gluon
degrees of freedom and thus reflect some deconfining aspects of a QCD
phase transition.

\subsubsection*{Pressure}

The pressure is obtained directly from the grand potential via
\begin{equation}
p(T,\mu_q) = - \Omega\left(T,\mu_q \right)
\end{equation}
with the vacuum normalization $p(0,0)=0$. In the model calculations it
is the primary observable and other thermodynamic quantities can be
obtained from it by differentiation. Note that this procedure differs
from the lattice approach, where for convenience the trace of the
energy-momentum tensor $\Theta^{\mu}_{\mu}$ is taken as the basis for
the thermodynamic observables evaluation~\cite{Bazavov:2009zn}. For
example, the pressure is then obtained by integration over the trace
of the energy-momentum tensor $\Theta^{\mu}_{\mu}$
~\cite{Karsch:2003jg}
\begin{equation}
p(T) = p(T_i) + \int_{T_i}^T dT' \frac{1}{T'^5} \Theta^{\mu}_{\mu}
\end{equation}
with the integration constant $p(T_i)$. In order to make the constant
as small as possible the temperature $T_i$ is adjusted accordingly,
e.g., $T_i \sim 100\MeV$~\cite{Karsch:2007dp} and $p(T_i)$ is finally
dropped. The suppression of this constant corresponds to an
approximation in the lattice data. 

Another normalization of the pressure lattice data consists in
applying the pressure from the hadron resonance gas (HRG),
$p_\text{HRG}(T_i)$, which provides a realistic description of the
pressure in the chirally broken phase.

Since $p_\text{HRG}(T_i=100\MeV)/T_i^4 \sim 0.8$
~\cite{Bazavov:2009zn} this normalization shifts the lattice data to
slightly higher values.

In \Fig{sfig:pressure} the pressure for the PQM and QM models,
normalized to the Stefan-Boltzmann limit (SB) of the PQM model, is
compared to lattice data. Explicitly, the pressure of $N_c^2-1$
massless gluonic and of $N_f$ massless fermionic degrees of freedom
tends to
\begin{equation}
\label{eq:SBpressure}
\frac{p_{SB}}{T^4} = (N_c^2-1)\frac{\pi^2}{45} + N_c N_f \frac{7\pi^2}{180}\ ,
\end{equation}
at high temperature, which we use for normalization. Of course, no
gluons are involved in the pure QM model and consequently, the first
term in \Eq{eq:SBpressure} does not contribute. Therefore, a lower
pressure is obtained in the QM model for high temperatures in the
figure. The same conclusion holds for the PQM with Fukushima's
potential. 

\begin{figure*}[tbp]
\centering
\subfigure[$\ $Energy density
$\epsilon/T^4$]{\label{sfig:epsilon}\includegraphics[width=\twofigs]{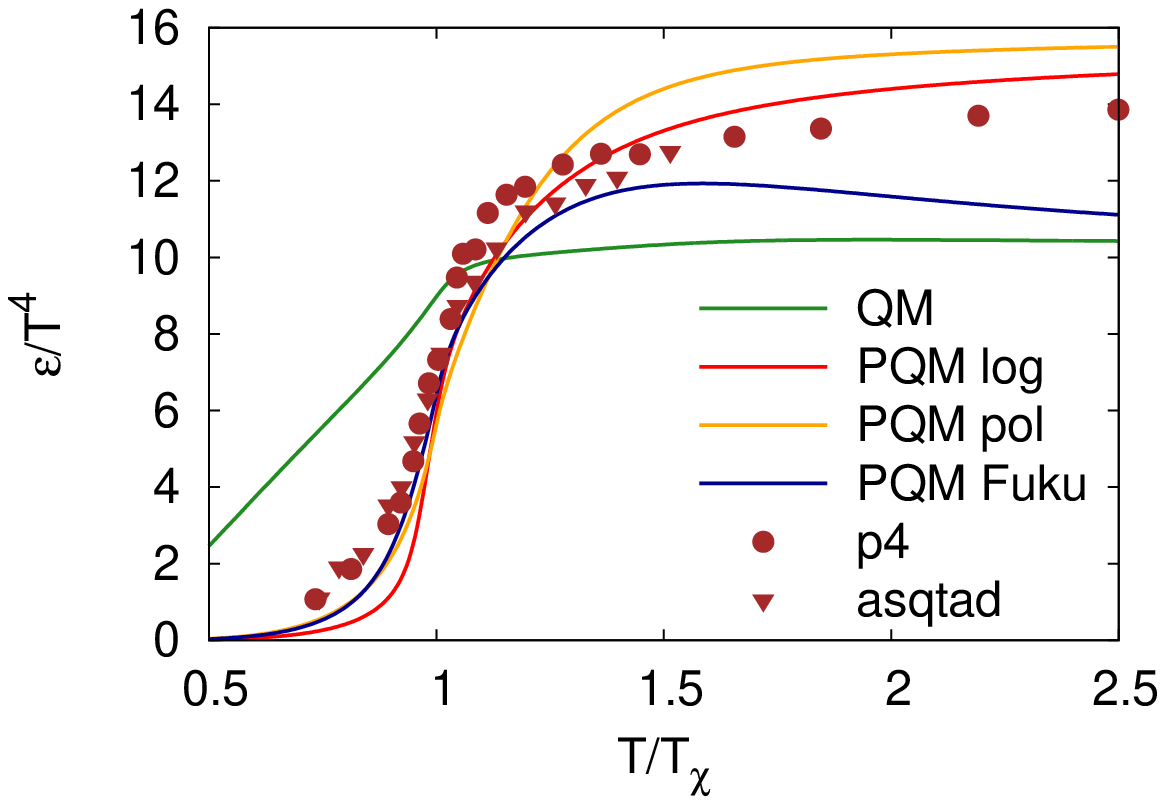}}
\subfigure[$\ $Entropy density
$s/T^3$]{\label{sfig:entropy}\includegraphics[width=\twofigs]{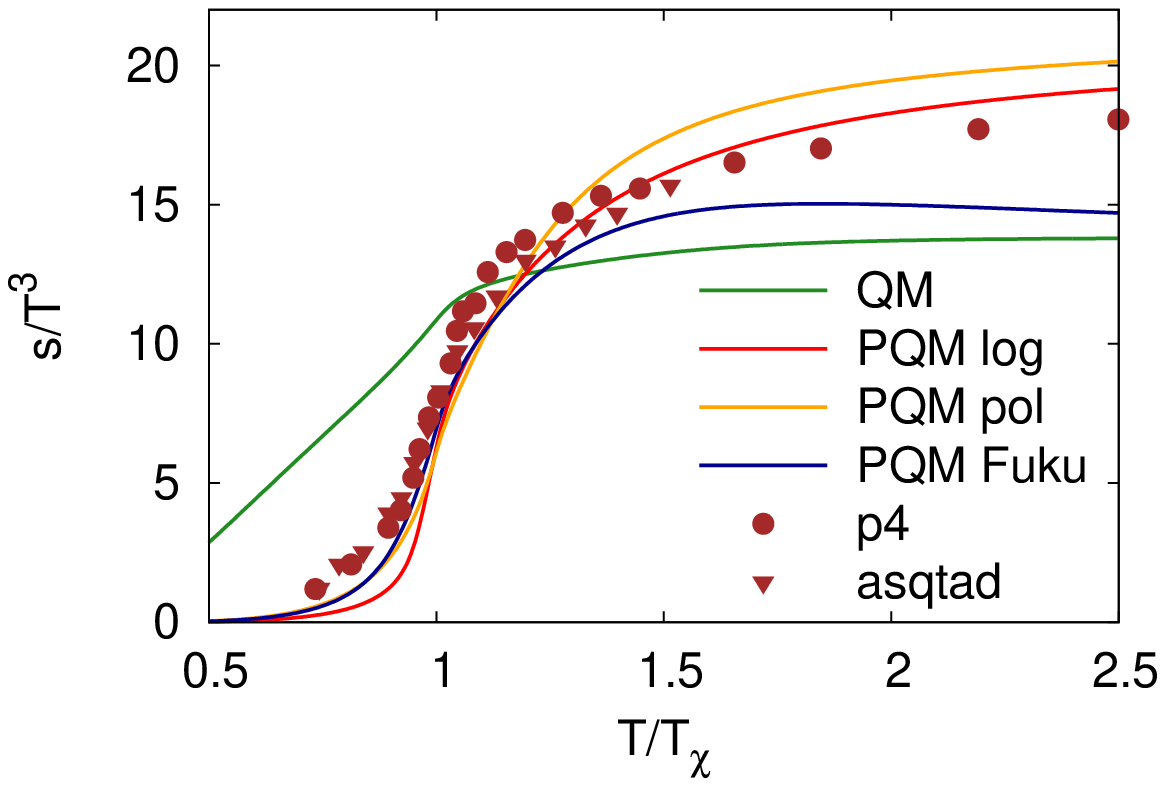}}
\caption{\label{fig:edensity}\coloronl The energy- (left panel) and
  entropy density (right panel) similar to Fig.~\ref{fig:pqmeos}.}
\end{figure*}

As expected, the QM model fails in describing the lattice data for all
temperatures, while all PQM model results are in reasonable agreement
with the data. 
The best agreement around $T_\chi$ with the lattice data is achieved
with Fukushima's potential but it fails for temperatures above
$1.5 T_\chi$ as already discussed in Sec.~\ref{sec:polypots}. Note
that for the HotQCD lattice data the transition temperature lies between
$T_\chi \approx 185-195\MeV$. We have used the average
value $T_\chi=190\MeV$ in the figures. Physical meson masses reduce the
pressure in particular around $T_\chi$ (dashed lines in
\Fig{sfig:pressure}).

The better agreement of the pressure of Polyakov-loop augmented chiral
models with lattice data has already been observed in two-flavor
studies such as in \cite{Schaefer:2007pw} for the PQM and in
\cite{Ratti:2005jh} for the PNJL model. However, these model findings,
obtained for physical pion masses, are compared with lattice data for
two rather heavy quark flavors~\cite{AliKhan:2001ek}. To some extent,
the agreement is therefore fortuitous since, on the one hand, mesonic
fluctuations are more suppressed in lattice simulations if heavier
masses are applied and, on the other hand, mean-field approximations
do not consider fluctuations around the mean-fields. In this work
larger pion masses are taken into account for the lattice comparison.

\subsubsection*{Interaction measure, energy density and
  entropy density}
The interaction measure $\Delta = e-3p$ is related to the grand
potential via a temperature derivative
\begin{equation}
\label{eq:e3p}
\frac{ \Delta}{T^4} = T \frac{\partial}{\partial
  T}{\left(\frac{p}{T^4}\right)} = -T \frac{\partial}{\partial
  T}\left( \frac{\Omega}{T^4} \right) 
\end{equation}
and can be obtained directly in lattice simulations from the trace of
the energy-momentum tensor
\begin{equation} 
  \frac{\Theta^{\mu}_{\mu}}{T^4}= \frac{\epsilon - 3p}{T^4} = \frac{
    \Delta}{T^4}\ .
\end{equation}
In the interaction measure finite volume and discretization effects
are more apparent than, e.g., in the pressure
(c.f.~\cite{Bazavov:2009zn}).

In \Fig{sfig:e3p} the model calculations for the interaction measure
are compared to $N_\tau=8$ lattice data. The height of the peak in the
lattice simulations depends significantly on the lattice size as well
as on the lattice action. Away from the phase transition the volume
dependence is less pronounced \cite{Bazavov:2009zn}. In the chirally
broken phase, the lattice data are close to all model curves. Around
the transition at $T \approx T_\chi$ Fukushima's potential is closest
to the asqtad-action data, while the remaining two model curves are in
better agreement with the p4-action data. In general, the peak height
decreases on larger lattices and in particular for the p4-action. For
these temperatures the asqtad-action shows a weak $N_\tau$-dependence
\cite{Bazavov:2009zn}. In contrast to the pressure, the logarithmic
and polynomial Polyakov model version cannot describe the lattice data
in the symmetric phase while Fukushima's ansatz approaches the data at
least up to $T\sim1.5 \Tl$.

In a PNJL study similar findings are obtained in the broken phase, but a
more rapid decrease is seen in the restored phase~\cite{Fukushima:2008wg}.
For physical pion masses, the interaction measure and the height of the
maximum decreases for temperatures below the transition. Above the
transition the interaction measure increases for smaller pion
masses.

To complete the thermodynamic comparison we also show the energy
density $\epsilon$ and the entropy density $s$, defined as
\begin{equation}
\label{eq:energydensity}
\epsilon = -p + T s \quad , \quad s = - \frac{\partial
  \Omega}{\partial T} \ ,
\end{equation}
in \Fig{fig:edensity} as a function of temperature. The rapid increase
of the energy density signals the liberation of light quark degrees of
freedom. It increases almost to the value of an ideal gas of massless
quarks and gluons.

\subsubsection*{Speed of sound $c_s$ and specific heat $c_V$}
For hydrodynamical investigations of relativistic heavy-ion collisions
the speed of sound $c_s$ is an important quantity. Its square at
constant entropy is defined by
\begin{equation}
  c_s^2 = \left.\frac{\partial  p}{\partial \epsilon}\right|_S =
  \left.\frac{\partial p}{\partial T}\right|_V \left/
    \left.\frac{\partial \epsilon}{\partial T}\right|_V = \frac{
      s}{c_V} \right., 
\end{equation}
where 
\begin{equation} 
\label{eq:cv}
c_V = \left.\frac{\partial \epsilon}{\partial T}\right|_V = -T \left.
  \frac{\partial^2 \Omega}{\partial T^2} \right|_V 
\end{equation}
denotes the specific heat capacity at constant volume. At a
second-order phase transition, where the specific heat diverges with
the critical exponent $\alpha$, the sound velocity drops to zero since
the entropy density stays finite, see e.g. \cite{Schaefer:1999em} for
a two-flavor QM model calculation.

\begin{figure}[t!]
  \centering \subfigure[$\ $Specific heat capacity $c_V/T^3$] {\label{sfig:cv}
    \includegraphics[width=\onefig]{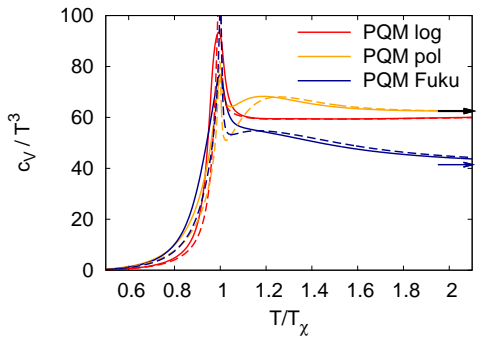}}
  \caption{\label{fig:cv}\coloronl The specific heat capacity in the
    PQM model as a function of temperature. Solid lines have been
    obtained for $m_\pi=220\MeV$, dashed lines for physical values.
    The arrows indicate the SB-limit. See text for details.}
\end{figure}

The specific heat calculated in the three-flavor PQM model at $\mu_q=0$
as a function of the temperature is displayed in \Fig{sfig:cv} for
three different Polyakov loop potentials. The solid lines in this figure
correspond to masses used in lattice simulations and the dashed lines to
physical masses. The specific heat grows with increasing temperatures,
peaks at the transition temperature and approaches the corresponding
SB-limit for high temperatures. The SB limits deviate due to the
differences in the contributing gluon degrees of freedom. For
Fukushima's potentials the first term in \Eq{eq:SBpressure} is
neglected. For smaller masses the peak is sharper. Just after the peak
a broad bump around 1.2 $T/T_\chi$ is found for the polynomial and
Fukushima's potential (for the latter with physical masses). Then
$c_V$ goes gradually to the ideal gas value from above.

\begin{figure*}[thbp]
  \centering \subfigure[$\ $ $c_s^2$ (solid lines) and $p/\epsilon$
  (dashed lines) versus
  $T/T_\chi$]{\label{sfig:cs2}\includegraphics[width=\twofigs]{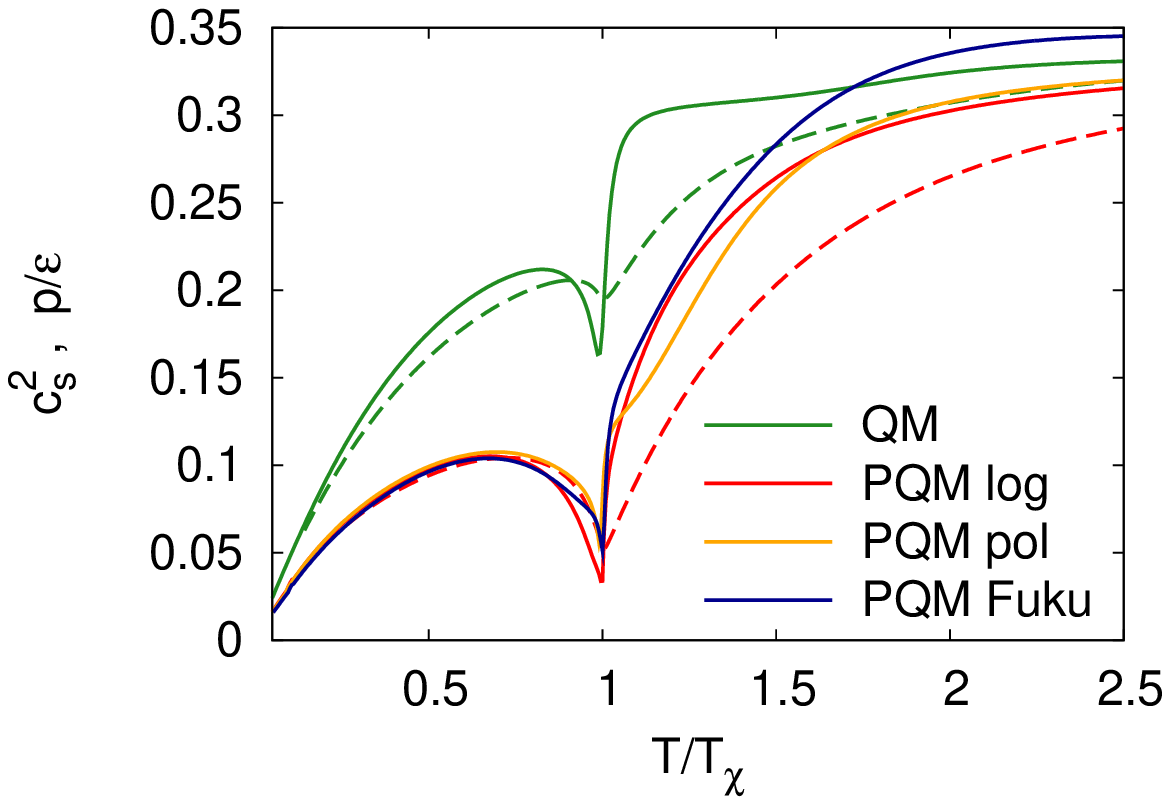}}
  \subfigure[$\ $ $c_s^2$ (solid lines) and $p/\epsilon$ (dashed
  lines) versus
  $\epsilon^{1/4}$]{\label{sfig:cs2lat}\includegraphics[width=\twofigs]{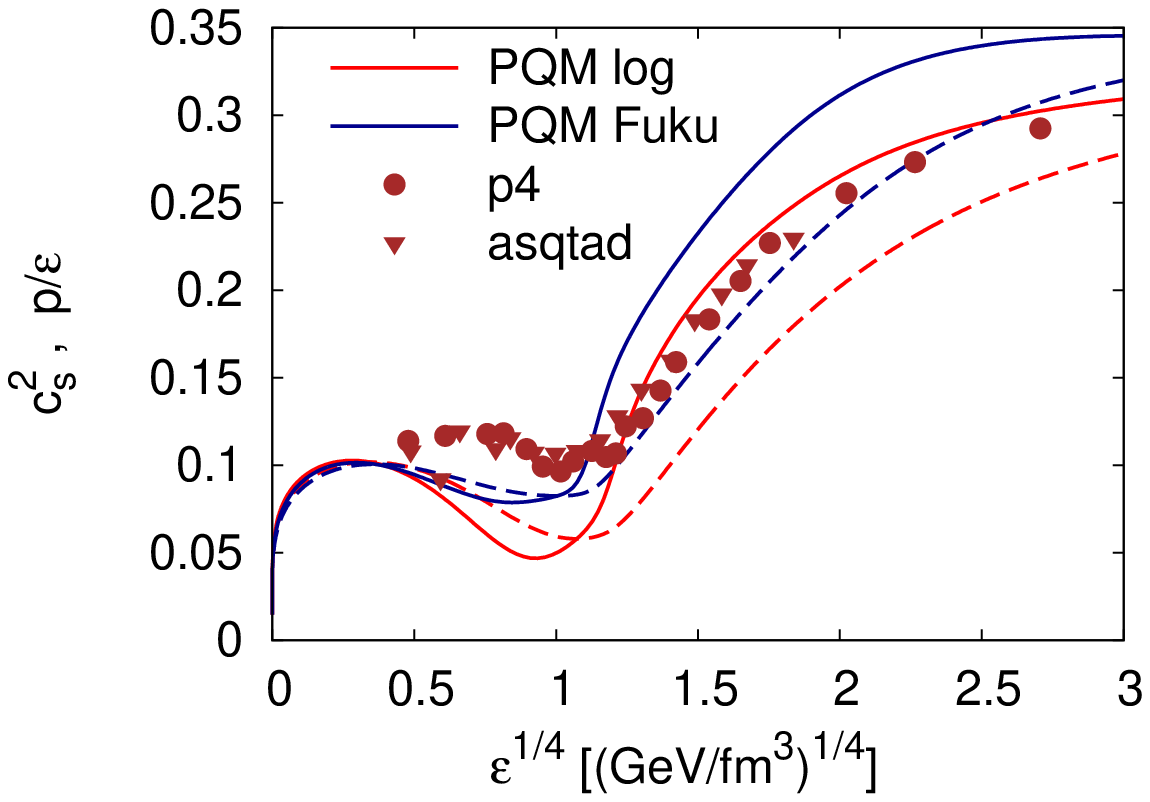}}
  \caption{\label{fig:pqmcs2}\coloronl Left panel: The squared speed
    of sound $c_s^2$ (solid lines) and the ratio $p/\epsilon$ (dashed
    lines) as a function of temperature. Right panel: For comparison
    with lattice data ($p/\epsilon$ from ~\cite{Bazavov:2009zn}) both
    quantities are also shown as a function of $\epsilon^{1/4}$.}
\end{figure*}

The velocity of sound as well as the ratio $p/\epsilon$ are shown in
\Fig{fig:pqmcs2} as a function of temperature (left panel) and
$\epsilon^{1/4}$ (right panel). The latter plot eliminates
uncertainties due to the normalization of the transition temperature.
For all models including the pure QM model $c_s^2$ approaches the
ideal gas value of $1/3$ at high temperatures of about 2.5$T_c$ and
can well be approximated by the ratio $p/\epsilon$. Furthermore, the
value of $p/\epsilon$ matches with the one of $c_s^2$ for low
temperatures. In between these two limits $c_s^2$ is always larger
than $p/\epsilon$ except near $T_\chi$, similar as in a two-flavor PNJL
analysis \cite{Ghosh:2006qh}.

Near $T_\chi$ it drops to a minimum of about 0.04 for the PQM model,
whereas the minimum is close to 0.16 if the Polyakov loop is
neglected. While the latter value 0.16 is close to the results for
pure YM theory on the lattice \cite{Gavai:2005da}
we find a minimum including the Polyakov loop almost half of the
softest point $P/\epsilon = 0.075$ in~\cite{Ejiri:2005uv} for
two-flavor staggered fermions. It is interesting that in our work
even at temperatures around $0.5 T_\chi$ $c_s^2$ is not larger than
0.1 if the Polyakov loop is included. This is in contrast to
Ref.~\cite{Mohanty:2003va} where a confinement model has been used and
values of about $c_s^2 \sim 0.2$ around $0.5 T_\chi$ and $c_s^2 =0.15$
around $T_\chi$ are found which correspond to our values for the pure
QM model.

Since the speed of sound is related to the interaction measure via
\begin{equation} 
  \frac{ \Delta}{\epsilon}  = \frac{ \epsilon -3p}{\epsilon }
  \approx 1 - 3 c_s^2\ ,  
\end{equation} 
a minimum in the speed of sound as expected near a phase transition
translates to a maximum in the conformal measure $\Delta/\epsilon$.
For high temperatures where $c_s^2$ reaches the SB limit of 1/3 the
conformal measure should vanish.

In the left panel of \Fig{fig:pqmcs2} where both quantities $c_s^2$
and $p/\epsilon $ are shown as a function of $\epsilon^{1/4}$ the best
agreement with the lattice data is achieved with Fukushima's potential
model.

\subsubsection*{Quark number susceptibilities and Kurtosis}

\begin{figure*}[t]
  \centering \subfigure[$\ $Light (solid) and strange (dotted) quark
  number susceptibilities $\chi_{l,s}$]{\label{sfig:suscepmu0ls}
    \includegraphics[width=\twofigs]{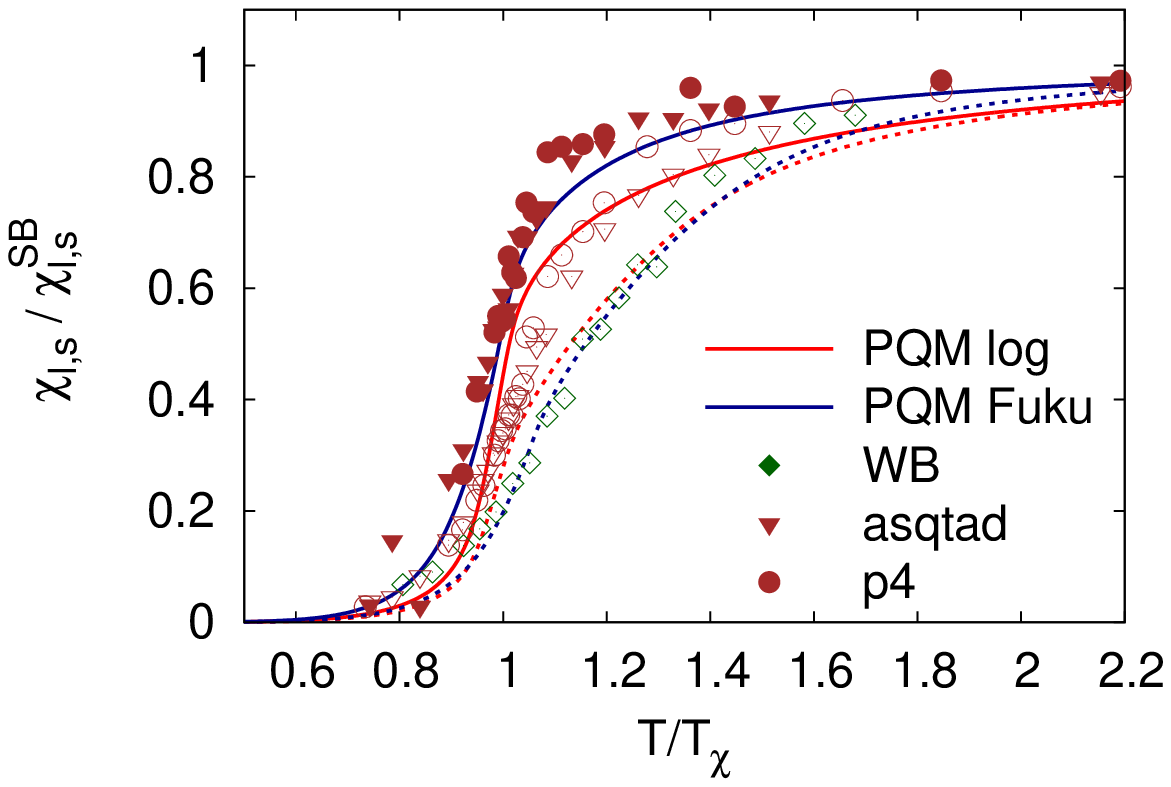}}
  \subfigure[$\ $Ratio of light and strange quark number susceptibilities
  $(2 \chi_s)/\chi_l$]{\label{sfig:suscepmu0lsratio}
    \includegraphics[width=\twofigs]{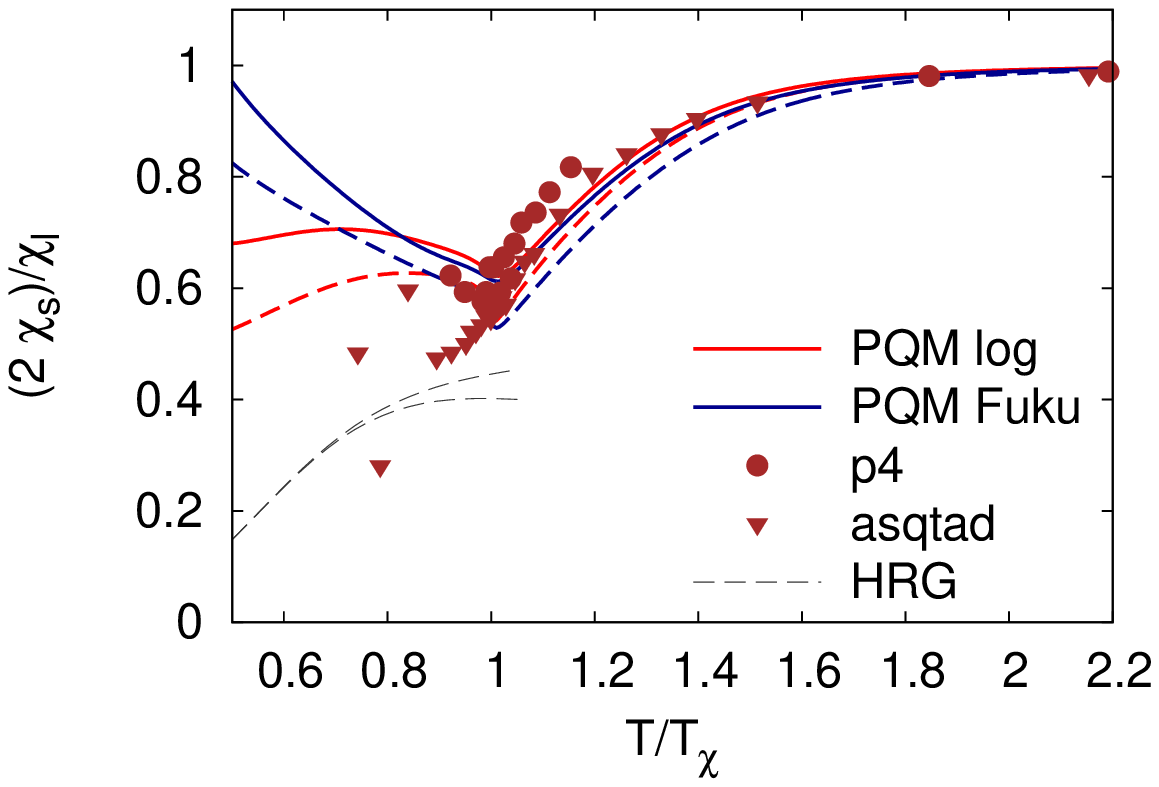}}
  \caption{\label{fig:suscepls}\coloronl Left panel: The light (solid
    lines/filled symbols) and strange (dotted lines/open symbols)
    quark number susceptibilities in comparison with lattice data
    ($N_\tau=8$) with larger lattice pion and kaon
    masses~\cite{Bazavov:2009zn}. For $\chi_s$ we also show results of
    the Wuppertal-Budapest group (WB) obtained with physical pion
    masses on $N_\tau=10$ lattices~\cite{Aoki:2009sc}. Right panel:
    The ratio of the strange and light quark number susceptibilities
    with physical meson masses (dashed lines) and lattice masses
    (solid lines) in comparison with lattice data. Also shown are the
    results for a hadron resonance gas model (HRG) including
    resonances up to 1.5 GeV (upper curve) and 2.5 GeV (lower curve)
    from~\cite{Bazavov:2009zn}.}
\end{figure*}

Quark number susceptibilities incorporate information on thermal
fluctuations of the degrees of freedom that carry a net number of
light or strange quarks. They are also related to the second order
cumulant of an expansion in the net quark number
$\delta n_q = n_q - \vev {n_q}$, e.g.~\cite{Stephanov:2008qz}.

In the transition region these susceptibilities changes because the
carriers of charge, strangeness or baryon number are heavy hadrons in
the low temperature phase but much lighter quarks at high
temperatures. Thus, at low temperatures they describe the fluctuations
of hadrons carrying net light quark or strangeness quantum numbers.
For example, the light quark number susceptibility, $\chi_l$, receives
basically only contributions from the lightest hadrons, the pions, at
low temperatures, i.e. $\chi_l/T^2 \propto \exp (-m_\pi/T)$ where the
index $l$ refers to the light quark flavors. Correspondingly,
$\chi_s/T^2$ receives contributions only from the lightest hadron that
carries strangeness. For high temperatures and $N_f$ quark flavors
these susceptibilities approach the SB value for an ideal massless
quark gas with $N_c$ colors, i.e. $\chi_q/T^2 \to N_f N_c/3$.

Furthermore, the quark number susceptibility $\chi_q$ corresponds to
the first Taylor coefficient $c_2$ of an expansion of the
thermodynamic pressure in powers of $\mu_q/T$
\begin{equation}
  \frac{ p}{T^4} \!=\! \sum_{n=0}^\infty
 c_n(T) \left( \frac{ \mu_q}{T}\right)^n \ 
 \text{with} \ 
c_n(T) \!=\! \frac{1}{n!}\left.\frac{\partial^n
    (p/T^4)}{\partial(\mu_q/T)^n}\right|_{\mu_q=0}\!\!\!\!\!.
\end{equation}

\begin{figure*}[t]
  \centering \subfigure[$\ $Kurtosis $R_{q}$]
  {\label{sfig:kurtosis}
    \includegraphics[width=\twofigs]{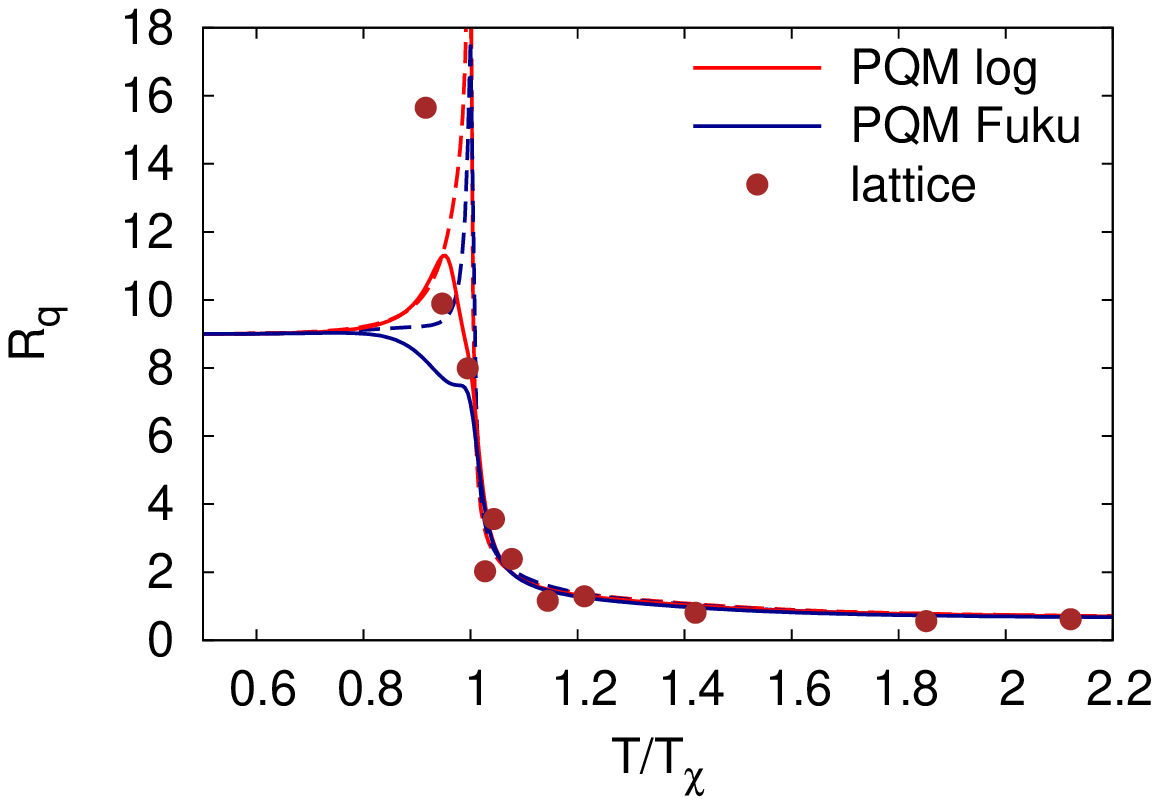}}
  \subfigure[$\ $Strange-quark kurtosis $R_{s}$]
  {\label{sfig:skurtosis}
    \includegraphics[width=\twofigs]{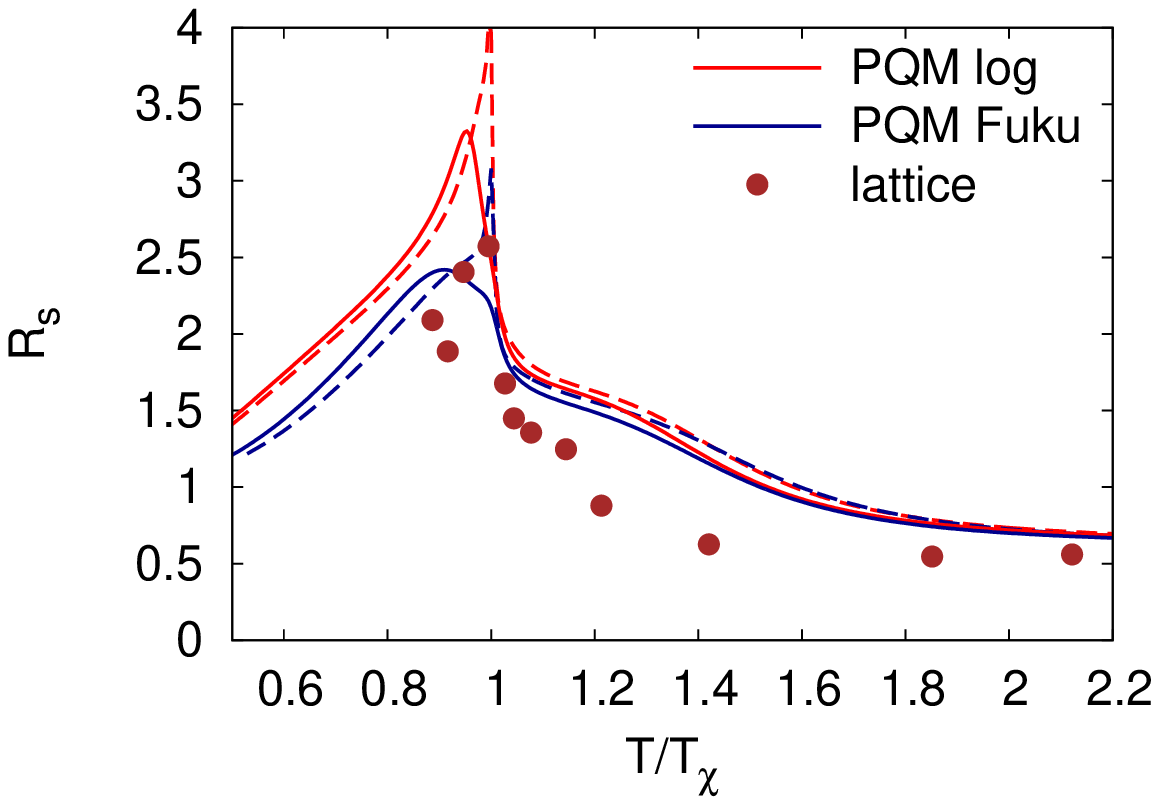}}

  \caption{\label{fig:kurtosis}\coloronl Left panel: Kurtosis $R_q$
    for uniform quark chemical potentials as a function of
    temperature.
    Right panel: Kurtosis $R_s$ for the strange quark sector. The
    lattice data ($N_\tau=6$) are taken from~\cite{Cheng:2008zh}.}
\end{figure*}

It can be obtained from the grand potential as follows
\begin{equation}
\label{eq:chiq}
\frac{\chi_q}{T^2} = 2 c_2 = \frac{ 1}{T^3V}\vev{(\delta n_q)^2} = -
\frac{\partial^2 \Omega}{T^2\partial \mu_q^2}   \ .
\end{equation}
At a second order transition $\chi_q$ diverges and, since the quark
number density stays finite, the isothermal compressibility also
diverges, see e.g.~\cite{Schaefer:2006ds}.

In the present (2+1)-flavor analysis we investigate the light and
strange quark number susceptibilities separately
\begin{equation}
\label{eq:chils}
\chi_l= - \frac{\partial^2 \Omega}{\partial \mu_l^2}\ , \quad \chi_s=
- \frac{\partial^2 \Omega}{\partial \mu_s^2}\ .
\end{equation}  

The required derivatives of the thermodynamic potential can be
calculated with a novel algorithmic differentiation technique,
described in \cite{Wagner:2009pm}. 

Both susceptibilities, normalized according to their SB limit, are
shown in \Fig{sfig:suscepmu0ls}. Around the chiral transition
temperature the light susceptibility $\chi_l$ raises faster than the
strange susceptibility. In general, the PQM model results for the
susceptibilities are below the lattice data. Closest to the model
findings are the Wuppertal-Budapest data for $\chi_s$ which have been
obtained with physical pion and kaon masses. However, the influence of
the pion mass in the strange sector is weak. The better agreement
might be related to the larger lattices ($N_\tau=10$) of the
Wuppertal-Budapest group.

To avoid unknown normalization factors in the lattice data
(cf.~\cite{Cheng:2008zh}) it is more convenient to consider the ratio
of the strange and light quark number susceptibilities as shown in
\Fig{sfig:suscepmu0lsratio}. In the restored phase and around the
transition the agreement with the lattice data is good. Note that the
two different lattice actions deviate strongly.

The ratio of the strange and light susceptibilities (right panel)
shows strong deviations of the model calculation from either the
lattice data or a HRG model calculation in the broken phase. This
might be attributed to the omission of mesonic fluctuations in the
mean-field approximation of the PQM model. Since the light and strange
fluctuations are mainly driven by the light pions and heavy kaons
respectively, this results in rather small ratios for the HRG model
calculation whereas values in between $0.5-1.0$ are found for the PQM
model.

Another quantity which probes deconfinement is the kurtosis of the
quark number fluctuations $R_{q}$. It is defined as the ratio of
the fourth derivative of the grand potential with respect to $\mu_q$,
i.e. $\chi_{q,4} \equiv - \partial^4 \Omega / \partial \mu_q^4$, and
the quark number susceptibility
\begin{equation}
R_{q}\equiv T^2\frac{\chi_{q,4}}{\chi_{q}} \ .
\end{equation}
It is sensitive to the net quark content of baryon number carrying
effective degrees of freedom. At small temperature, the effective
degrees of freedom in the PQM model are three-quark states since the
Polyakov loop vanishes there and suppresses the one- and two-quark
states in the thermodynamic potential, see \Eq{eq:Omegaqq}. Thus, the
quark/antiquark contribution of the thermodynamic potential is
basically that of a non-interacting system with particles and
antiparticles carrying baryon number $B=\pm 1$. Consequently, the
kurtosis tends in the low-temperature confined phase with $|B|=1$ to
$R_{q} = (N_c B)^2 = 9$. Without the Polyakov loop, e.g. in a pure QM
model, the kurtosis would tend to one since in this case single quarks
with $|B|=1/3$ would contribute at all temperatures. Thus, the correct
low-temperature value for the kurtosis is obtained only with the
Polyakov loop.

At high temperatures the Polyakov loop tends to one,
\Fig{sfig:pqmpoly}. As a consequence, the quark/antiquark contribution
of the thermodynamic potential, \Eq{eq:Omegaqq}, corresponds to a
system of non-interacting quark gas and the kurtosis reaches the SB
value, $R_{q} \rightarrow 6 /\pi^2$. Both temperature limits of $R_q$
are precisely reproduced with the PQM model as shown in
\Fig{fig:kurtosis}. The kurtosis is almost temperature independent far
below and above the transition. But around the transition it is
strongly dependent on the meson masses and the effective
implementation of the Polyakov loop and its coupling to the quark
sector of the model. The PQM model calculation is consistent with
lattice data ($N_\tau=6$) \cite{Cheng:2008zh}. At the transition the
lattice data depict a peak behavior in $R_q$ for larger meson masses
($m_\pi\sim220\MeV$) but the volume dependence and errors there are
still large. As already mentioned, only the PQM model provides the
correct low-temperature limit $R_q=9$ and reaches the high-temperature
limit already at $T\sim 1.3T_\chi$ similar to a two-flavor PQM
analysis with a polynomial Polyakov loop potential
\cite{Stokic:2008jh}. For Fukushima's potential we observe a similar
dependence of the peak structure on the pion mass, i.e., it vanishes
for larger masses. The peak structure is more pronounced for a
logarithmic potential and survives if the masses are increased.
Interestingly, in a three-flavor PNJL analysis with Fukushima's
potential no peak is seen even for physical pion masses
\cite{Fukushima:2008wg}. A two-flavor PNJL model with a logarithmic
potential gives a temperature dependent kurtosis in the broken phase
for $T<T_\chi$, which we also find in this work if we consider only
the kurtosis for light flavors, i.e., $R_{l}=
T^2{\chi_{l,4}}/{\chi_{l}}$.

In Fig.~\ref{sfig:skurtosis} the kurtosis for the pure strange sector
$ R_{s}= T^2 {\chi_{s,4}}/{\chi_{s}} $ is presented. For the lattice
simulations $N_\tau=6$ and physical strange quark masses have been
used. Note that chiral symmetry in the strange sector is restored at
$\Ts \sim 1.3 \Tchi$. For these temperatures the lattice data are
already quite close to the SB-limit while the model analysis still
yields an enhanced ratio. The strange quark kurtosis rises, and peaks
at the chiral symmetry restoration for the light quarks. It stays
above the lattice data until the symmetry is fully restored also in the
strange sector and drops finally to its SB limit for high
temperatures.

\section{summary}
\label{sec:summary}
In the present work we have extended a previously introduced
two-flavor Polyakov-quark-meson (PQM) model~\cite{Schaefer:2007pw} to
three quark flavors. This model with the Polyakov loop can address
certain confinement issues and is an improvement of the known chiral
quark-meson model with three quark flavors. Several aspects, in
particular the bulk thermodynamics, of the ($2+1$)-flavor QCD phase
transitions are in much better agreement with recent ($2+1$)-flavor
lattice simulations.

The Polyakov loop potential is not uniquely determined and different
versions are available. Here, we have considered three different
effective Polyakov loop potentials which all reproduce a first-order
phase transition in the pure gauge sector of the theory at $T_c = 270$
MeV. In two Polyakov potential versions a further parameter $T_0$
enters which we have varied in order to investigate the influence of
finite quark masses on the transition. The scalar and pseudoscalar
chiral sector of the PQM model is fitted to the mass spectrum and
decay constants in the vacuum. The input masses are experimentally
well-known except for the scalar $\sigma$-resonance. Different sigma
mass values have a strong influence on the chiral phase transition. In
this work we have additionally considered various input values for the
$\sigma$ mass similar as in \cite{Schaefer:2008hk}.

For some parameter combinations a coincidence of the chiral and
deconfinement phase transition at vanishing chemical potential is
found which we have used to study the bulk thermodynamics of the
finite temperature transition.

The model results are confronted with recent lattice simulations for
$2+1$ quark flavors of the RBC-Bielefeld~\cite{Cheng:2008zh} and
HotQCD~\cite{Bazavov:2009zn} collaborations. The latter lattice group
uses lattices with temporal extent $N_\tau=8$ and the quark masses in
these simulations are still above the expected physical ones. These
quark masses correspond to a kaon mass of $m_K=503 \MeV$ and a pion
mass of $m_\pi=220 \MeV$. This difference in the masses influences the
bulk thermodynamics and its effect is also addressed in this work.

In addition, for the chiral order parameter and the strange quark
number susceptibility lattice results for physical quark masses on
$N_\tau = 10$ lattices of the Wuppertal-Budapest group
\cite{Aoki:2009sc} are available and compared with the model analysis.

We observe a nice matching with the lattice data for the chiral
transition, although the transition is smoother on the lattice, in
particularly for the Wuppertal-Budapest results. The lattice data for
the Polyakov loop are below the model data, especially in the
deconfined phase.

A very good agreement of the lattice data with the model is found for
the equation of state in the transition region and for temperatures up
to $1.5 \Tchi$.

Due to the omission of mesonic fluctuations in the mean-field
approximation a deviation of the model and the lattice data is
expected at low temperatures. Similarly, at high temperatures, the
gluon dynamics is not fully addressed in the model which may also
cause further deviation.

The pressure is not very sensitive to different implementations of the
Polyakov loop potential and to details of the lattice simulations.
This is in contrast to the interaction measure where Fukushima's
potential describes best the lattice data in this temperature regime,
especially, with regard to the asqtad-action.

For higher derivatives of the thermodynamic potential such as the
quark-number susceptibilities the picture changes. The model results
are always below the lattice data. This is the case not only in the
chirally broken phase which is understandable due to the omission of
mesonic fluctuations but also in the symmetric phase.

Furthermore, there are also some normalization factors involved in the
lattice data as well as some volume dependence. For example, the
susceptibilities decrease from $N_\tau=4$ to $N_\tau=8$ lattices.
Interestingly, the $N_\tau=10$ data for $\chi_s$ of the
Wuppertal-Budapest group are very close to our model results. However,
the normalization factors drop out in the ratio $\chi_s/\chi_l$. As a
consequence the model results come closer to the lattice in the
symmetric phase. In the broken phase the PQM model is far off the
lattice and HRG results which might be an artifact of the mean-field
approximation.

In summary, we found that the PQM model gives a reasonable description
of the bulk thermodynamics in the QCD transition region. The current
implementation of the Polyakov loop into the chiral quark-meson model
is certainly an improvement but still further developments are
required. It seems that the remnant of the first-order transition in
the pure gauge sector modifies the chiral and deconfinement transition
at finite quark masses such that the transition becomes much sharper
compared to lattice data. The sudden liberation of the fermionic
degrees of freedom in the model triggers this sharp behavior. This
might change if one goes beyond the mean-field approximation. In this
case the mesonic fluctuations also contribute to the thermodynamics
and drive the phase transition.

\subsection*{Acknowledgments}
We thank F. Karsch, C. Schmidt and Z. Fodor for providing lattice data
and discussions. The work of MW was supported by the Alliance Program
of the Helmholtz Association (HA216/ EMMI) and BMBF grants 06DA123 and
06DA9047I. JW was supported in part by the Helmholtz International
Center for FAIR. We further acknowledge the support of the European Community-Research
Infrastructure Integrating Activity
Study of Strongly Interacting Matter under the Seventh Framework Programme of EU.


\end{document}